\documentclass[prl,twocolumn,showpacs,floatfix,amsfonts]{revtex4}
\usepackage{graphicx,graphics,color,epsfig}% Include figure files
\usepackage{bm}
\usepackage{amsmath}
\usepackage{amssymb}
\usepackage{epstopdf}
\begin{document}
\preprint{}
\title{Restoration of  $hc/2e$ Magnetic Flux Periodicity  in a Hollow  $d$-Wave Superconducting Cylinder}
\author{Jian-Xin Zhu}
\email{jxzhu@lanl.gov}
\homepage{http://theory.lanl.gov}
\affiliation{Theoretical Division, Los Alamos National Laboratory,
Los Alamos, New Mexico 87545, USA}

\begin{abstract}

The magnetic flux dependence of order parameter and supercurrent is studied  in a hollow $d$-wave superconducting cylinder. It is shown that the existence of line nodal quasiparticles in a pure $d_{x^2-y^2}$ pairing state gives rise to an $hc/e$ periodicity in the order parameter and a first-order quantum phase transition for a large system size.  We demonstrate that the flux periodicity in the supercurrent is sensitive to the detailed electronic band structure and electron filling factor. In particular, we find that, in cooperation with the increase of the cylinder circumference,  the $hc/2e$ periodicity can be restored significantly  in the supercurrent by avoiding the particle-hole symmetry point. 
A similar study of a $d_{x^2-y^2}+id_{xy}$ pairing state verifies the peculiarity of unconventional superconductors with nodal structure.
\end{abstract}
\maketitle

%\narrowtext

A fundamental property of all known superconductors is the formation of Cooper pairs~\cite{BCS:1957} in the superconducting state. A far-reaching implication of this fact is the quantization of magnetic flux in units of $hc/2e$ in multiply connected superconducting geometries. The $hc/2e$ flux quantization has been used as a proof of electron pairing nature of 
both conventional~\cite{BSDeaver:1961,RDoll:1961,NByers:1961,WBrenig:1961} 
and high-temperature~\cite{CEGough:1987} superconductors  in the superconducting state.
Other related phenomena include the quantum oscillation in the transition 
temperature~\cite{WALittle:1962} and magnetic vortices each carrying an $hc/2e$ flux quantum~\cite{AAAbrikosov:1957,UEssmann:1967}. 

Quantum mechanically, there is no fundamental reason why the minimal flux periodicity must be $hc/2e$ in a superconductor. The gauge invariance can only guarantee a fundamental period of $\Phi_0=hc/e$~\cite{NByers:1961,FBloch:1970}. Only when all Cooper pairs move in the same group velocity, can a substantial $\Phi_0/2$ periodicity be obtained.  The  $\Phi_0$-periodicity has been in mesoscopic conventional superconducting  rings~\cite{JXZhu:1994,KCzajka:2005,TCWei:2007,VVakaryuk:2008} due to the level discreteness and Landau depairing effect. More surprisingly, recent studies have shown a severe breaking of $hc/2e$-periodicity in a 
$d$-wave superconducting loop~\cite{FLoder:2008,YuSBarash:2008,VJuricic:2008}, as 
a result of the Cooper-pair angular 
momentum selection for the existence of Doppler-shifted zero-energy states.
%, a phenomenon robust against the changes in geometry, the increase of loop arms, and scattering from static impurities~\cite{FLoder:2008}.
%Within the Andreev equation analysis, the interesting results were interpreted by the angular 
%momentum selection for the existence of Doppler-shifted zero-energy states~\cite{YuSBarash:2008}.
%The interesting results later on interpreted by the existence of Doppler-shifted zero-energy states only for pairings with even angular momenta of the center of mass of Cooper pairs, within the Andreev equation analysis~\cite{YuSBarash:2008}.

In this Letter, by  a systematic study of the supercurrent in a hollow $d$-wave superconducting cylinder,
we show an $hc/e$ magnetic periodicity in the $d$-wave order parameter and demonstrate that the flux periodicity in the supercurrent is sensitive to the detailed electronic band structure and electron filling factor. In particular, we find that the breaking of $hc/2e$ periodicity in the case of $d$-wave pairing is closely related to the particle-hole symmetry in the normal state band structure, which gives rise to the van Hove singularity. When the particle-hole symmetry point is avoided, the $hc/2e$ periodicity can be restored almost completely in the supercurrent.

\begin{figure}[t]
%%\centerline{\psfig{fig1.eps.eps,width=8cm}}
\includegraphics[width=8cm, angle=0]{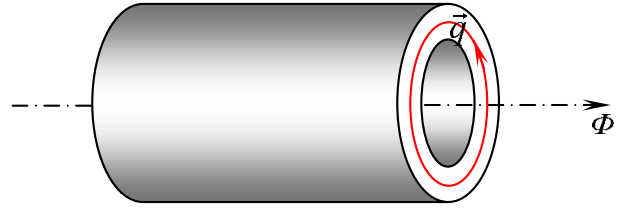}
\caption{(Color online)  Schematic drawing of a hollow $d$-wave superconducting cyliner. A magnetic flux threads the cylinder parallel to its axis.
}
\label{FIG:Cylinder}
\end{figure}

To be specific, we consider a hollow $d$-wave superconducting cylinder, as schematically shown in Fig.~\ref{FIG:Cylinder}.  Experimentally, the cylinder can be formed by a high-temperature superconductor film with its normal perpendicular to the CuO$_2$ plane.  A magnetic flux $\Phi$ threads the cylinder parallel to its axis, and also to the crystal $b$-axis of the CuO$_2$ plane. 
 Due to the weak interlayer coupling, the problem can be reduced to one on an essentially two-dimensional system.  We define the $x$- and $y$-axis to be perpendicular and parallel to the flux direction, respectively. This set up can also avoid the nucleation of Abrikosov vortices, which will complicate the analysis. By choosing a gauge, where the vector potential does not appear explicitly in the Hamiltonian, the Bogoliubov-de Gennes (BdG) equations can be written as~\cite{PGdeGennes:1965}:
 \begin{equation}
 \sum_{j} \left( 
 \begin{array}{cc}
 \mathcal{H}_{ij}    & \Delta_{ij} \\
 \Delta_{ij}^{*}   & -\mathcal{H}_{ij}^{*} 
 \end{array} \right) 
 \left( \begin{array}{c}
 u_{j}^n \\ v_{j}^n 
 \end{array} 
 \right) = E_{n}  
 \left( \begin{array}{c}
 u_{i}^n \\ v_{i}^n 
 \end{array} 
 \right)\;,
 \label{EQ:BdG}
 \end{equation}
subject to the flux-modified boundary condition
\begin{equation}
\left(
\begin{array}{c}
u_{ix+Nx,iy}^n \\
v_{ix+Nx,iy}^n
\end{array}
\right) = \left( \begin{array}{cc}
e^{i2\pi\Phi/\Phi_0} & 0 \\
0  & e^{-i2\pi\Phi/\Phi_0} 
\end{array} \right)
\left(
\begin{array}{c}
u_{i}^n \\
v_{i}^n 
\end{array} \right)\;,
\label{EQ:MBC}
\end{equation}
Here $(u_i^n,v_i^n)$ are the eigenfunctions corresponding to eigenvalues $E_n$.  The single particle Hamiltonian $\mathcal{H}_{ij} =
-t_{ij} -\mu \delta_{ij}$ with $t_{ij}$ and $\mu$ being the hopping integral and chemical potential. We consider the nearest neighbor, $t$, and next-nearest neighbor, $t^{\prime}$, hopping integral. The bond order parameter for $d$-wave pairing is determined self-consistently as $\Delta_{ij} = (V_{x^2-y^2}/4) \sum_{n} [u_{i}^{n}v_{j}^{n*} + u_{j}^{n}v_{i}^{n*}] 
\tanh(E_{n}/2k_{B}T)$ with $V_{x^2-y^2}$ being the pairing strength in the $d_{x^2-y^2}$ channel. Notice that the quasiparticle excitation energy is measured with respect to the Fermi energy.

If the electron wave vector is $\mathbf{k}=(k_x,k_y)$ and that for the collective drift motion (superfluid motion) of the paired electrons is $\mathbf{q}=(q_x,q_y)$, the initial $\mathbf{k}$ and $-\mathbf{k}$ pairing is adjusted to pair the states $\mathbf{k}+\mathbf{q}$ and $-\mathbf{k}+\mathbf{q}$.  The solution to the BdG equations is then found as:
\begin{equation}
\left( \begin{array}{c}
u_i \\ v_i 
\end{array} \right) = \left(
\begin{array}{cc}
e^{i(\mathbf{k}+\mathbf{q})\cdot \mathbf{r}_i} & 0 \\
0 & e^{i(\mathbf{k}-\mathbf{q})\cdot \mathbf{r}_i} 
\end{array} \right) 
\left(   \begin{array}{c}
u_{\mathbf{k},\mathbf{q}} \\
v_{\mathbf{k},\mathbf{q}} 
\end{array} \right) \;.
\end{equation}
Here, corresponding to $E^{(\pm)}_{\mathbf{k},\mathbf{q}} = Z_{\mathbf{k},\mathbf{q}} \pm E_{\mathbf{k},\mathbf{q}}^{(0)}$, the electron and hole components  of the quasiparticle amplitude are given by 
$(u_{\mathbf{k},\mathbf{q}}, v_{\mathbf{k},\mathbf{q}}) = (u_{\mathbf{k},\mathbf{q}}^{(0)}, v_{\mathbf{k},\mathbf{q}}^{(0)})$ and $(v_{\mathbf{k},\mathbf{q}}^{(0)}, -u_{\mathbf{k},\mathbf{q}}^{(0)})$
with
\begin{equation}
\vert u_{\mathbf{k},\mathbf{q}}^{(0)}\vert^{2} = \frac{1}{2} \biggl{(} 1 + \frac{Q_{\mathbf{k},\mathbf{q}}}{E^{(0)}_{\mathbf{k},\mathbf{q}}}\biggr{)}\;,\;\;\vert v_{\mathbf{k},\mathbf{q}}^{(0)}\vert^{2} = \frac{1}{2} \biggl{(} 1 - \frac{Q_{\mathbf{k},\mathbf{q}}}{E^{(0)}_{\mathbf{k},\mathbf{q}}}\biggr{)}\;.
\end{equation}
The quantities $Q_{\mathbf{k},\mathbf{q}} = [\xi_{\mathbf{k}+\mathbf{q}} + \xi_{\mathbf{k}-\mathbf{q}}]/2$, $Z_{\mathbf{k},\mathbf{q}} = [\xi_{\mathbf{k}+\mathbf{q}} - \xi_{\mathbf{k}-\mathbf{q}}]/2$, and 
$E^{(0)}_{\mathbf{k},\mathbf{q}} = [Q_{\mathbf{k},\mathbf{q}}^{2} + \Delta_{\mathbf{k}}^{2}]^{1/2}$.
In the tight-binding approximation, up to the next-nearest neighbor, the conduction electrons have the normal state  dispersion, $\xi_{\mathbf{k}} = -2t (\cos k_x  + \cos k_y ) -4t^{\prime} \cos k_x \cos k_y -\mu$. The $d$-wave superconducting gap dispersion is given by
 $\Delta_{\mathbf{k}} = 2\Delta_{d}\phi_{\mathbf{k}}$, where $\phi_{\mathbf{k}}= \cos k_x - \cos k_y$, and $\Delta_{d}(\mathbf{q})$ is determined self-consistently:
 \begin{eqnarray}
 \Delta_{d}(\mathbf{q}) &=& \frac{V_{x^2-y^2}}{8N_{L}}\sum_{\mathbf{k}} \frac{\phi_{\mathbf{k}}\Delta_{\mathbf{k}}}{E_{\mathbf{k},\mathbf{q}}^{(0)}} 
 \biggl{[}\tanh\biggl{(} \frac{E_{\mathbf{k},\mathbf{q}}^{(0)}+ Z_{\mathbf{k},\mathbf{q}}}{2k_{B}T}\biggr{)} 
 \nonumber \\  
 && + \tanh\biggl{(} \frac{E_{\mathbf{k},\mathbf{q}}^{(0)} - Z_{\mathbf{k},\mathbf{q}}}{2k_{B}T}\biggr{)}
 \biggr{]}\;,
 \end{eqnarray}
where $N_{L}=N_x \times N_y$ and the quantities $N_x$ and $N_y$ represent the circumference and length of the hollow cylinder. Rigorously, the bond order parameters along the $x$ and $y$ directions do not follow the relation $\Delta_{x}=-\Delta_{y}$ in the presence of magnetic flux. Here we have imposed the restriction of $\Delta_{x}=-\Delta_{y}$ to enforce a rigorous $d$-wave symmetry.  Notice that  $\Delta_{d}$ is now a function of $\Phi/\Phi_0$. From the boundary condition given by Eq.~(\ref{EQ:MBC}), one can find~\cite{JXZhu:1994} that the components of wave vectors $k_x= 2\pi (n_x-m/2)/N_x $,
$q_x= 2\pi (\Phi/\Phi_0 + m/2)/N_x $, while  $k_y= 2\pi n_y/N_y$,
and $q_y=0$, where $n_{x(y)}$ and $m$ are integers. In particular, $m$ is determined by minimizing $\vert q_x \vert$ for a given value of magnetic flux $\Phi$. The electron filling factor and 
the single nearest-neighbor bond current flowing around the cylinder are computed via,
\begin{equation}
n_e = \frac{2}{N_{L}} \sum_{\mathbf{k}} [f(E_{\mathbf{k},\mathbf{q}}^{(+)})\vert u_{\mathbf{k},\mathbf{q}}^{(0)}\vert^{2} + f(E_{\mathbf{k},\mathbf{q}}^{(-)})\vert v_{\mathbf{k},\mathbf{q}}^{(0)}\vert^2] 
\;,
\end{equation}
\begin{eqnarray}
I &=& \frac{4te}{N_{L}} \sum_{\mathbf{k}} [f(E_{\mathbf{k},\mathbf{q}}^{(+)})\vert u_{\mathbf{k},\mathbf{q}}^{(0)}\vert^{2} + f(E_{\mathbf{k},\mathbf{q}}^{(-)})\vert v_{\mathbf{k},\mathbf{q}}^{(0)}\vert^2] 
\nonumber \\
&&\times \sin(k_x + q_x)\;,
\end{eqnarray}
respectively.
%The momentum-dependent spectral density  is given by 
%\begin{eqnarray}
%A(\mathbf{k},E) &=& 2[\vert u_{\mathbf{k},\mathbf{q}}^{(0)}\vert^{2} 
%\delta(E-E^{(+)}_{\mathbf{k},\mathbf{q}}) \nonumber \\
% && + \vert v_{\mathbf{k},\mathbf{q}}^{(0)}\vert^{2} \delta(E-E^{(-)}_{\mathbf{k},\mathbf{q}})]\;,
%\end{eqnarray}
%while the density of states 
%$\rho(E) = \frac{1}{N_{L}} \sum_{\mathbf{k}} A(\mathbf{k},E) $.
A factor of 2 has been included to account for the spin degeneracy.

In the numerical calculations, we take $k_{B}=t=1$.  Throughout the work, the energy is measured in units of $t$ unless specified otherwise. The temperature is fixed at $T=0.01$ and the $d$-wave channel pairing interaction is chosen to be $V_{x^2-y^2}=4$. Both the hopping parameter $t^{\prime}$ and the electron filling $n_e$ will be changed. For a given $n_e$, the chemical potential should be 
adjusted and, therefore, will be a function of $\Phi/\Phi_0$.

\begin{figure}[t]
%%\centerline{\psfig{fig2.eps,width=8cm}}
\includegraphics[width=8cm, angle=0]{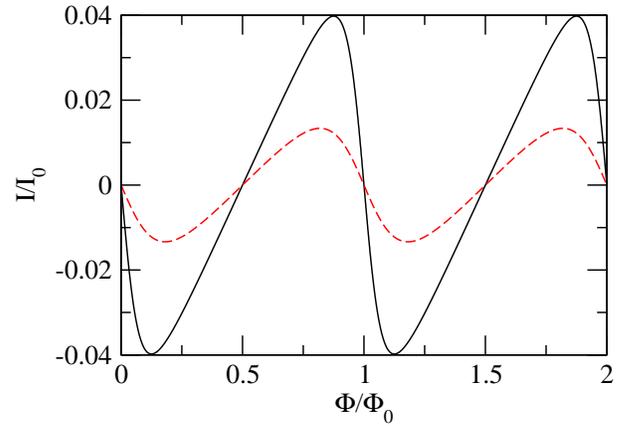}
\caption{(Color online)  Flux dependence of the persistent current for a normal state metallic cylinder of size $N_{L}=40^2$ (solid line) and $N_{L} = 80^2$ (dashed line), with $n_e=1.0$ and $t^{\prime}=0$. The current is measured in units of $I_0=et$. The other parameter values are defined in the main text. 
} \label{FIG:NormalCurrent}
\end{figure}

\begin{figure}[t]
%%\centerline{\psfig{filefig2.eps,width=8cm}}
\includegraphics[width=8cm, angle=0]{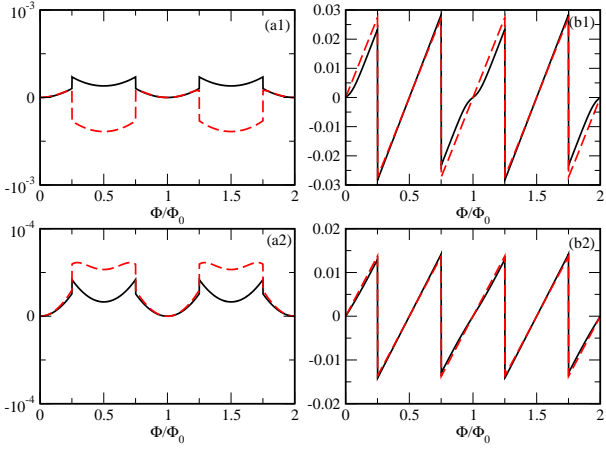}
\caption{(Color online)  Flux dependence of the $d$-wave order parameter (a1-a2) and persistent current (b1-b2) for a $d$-wave superconducting cylinder of zie $N_L=40^2$ (a1-b1) and $N_{L}=80^2$ (a2-b2), with $n_e=1.0$ (solid line) and $n_e=0.8$ (dashed line). In (a1-a2),  the relative amplitude 
of the $d$-wave right parameter, $[\Delta_{d}(\Phi) -\Delta_{d}(\Phi=0)]/\Delta_{d}(\Phi=0)$, is shown.
Here $t^{\prime}=0$ and the other parameter values are defined in the main text.
} \label{FIG:SuperCurrent1}
\end{figure}

In Fig.~\ref{FIG:NormalCurrent}, we show the flux dependence of the persistent current in a normal state metallic cylinder, where there is no existence of superconducting order parameter. These results are  known from the study of persistent current in normal state mesoscopic 
rings~\cite{MButtiker:1983} in the presence of an Aharonov-Bohm flux~\cite{YAharonov:1959}. The main point is that the persistent current in a normal state ring has a periodicity of $\Phi_0$. We present them here to demonstrate that our theoretical formulae designed for the superconducting state can reduce to describe the normal state, and
to provide a starting point for our discussion on the case of $d$-wave superconducting state below.

\begin{figure}[t]
%%\centerline{\psfig{filefig2.eps,width=8cm}}
\includegraphics[width=8cm, angle=0]{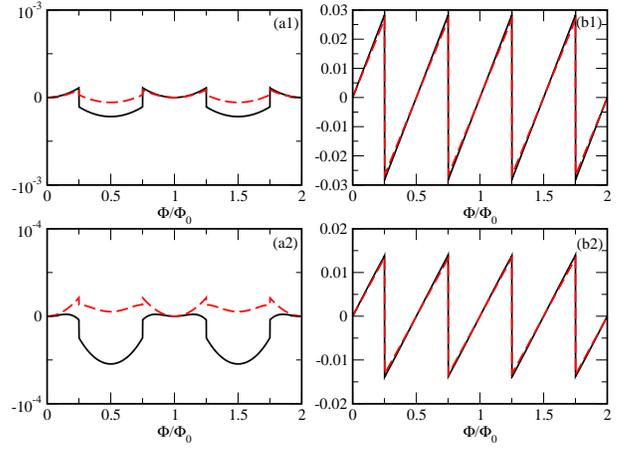}
\caption{(Color online)  The Same as Fig.~\ref{FIG:SuperCurrent1} but with $t^{\prime}=-0.2$.
} \label{FIG:SuperCurrent2}
\end{figure}

Figure~\ref{FIG:SuperCurrent1} shows the flux dependence of the 
$d$-wave order parameter and persistent current for various electron filling and system size but with $t^{\prime}=0$.  Due to the existence of nodal quasiparticle in a $d_{x^2-y^2}$ pairing symmetry, the order parameter oscillates in the magnetic flux with a period of $\Phi_0$, even for larger cylinders, although the oscillation amplitude is even smaller. The oscillation pattern is strongly sensitive to the system size. In particular, it is seen that the system experiences a first-order quantum phase transition when the flux across $\Phi/\Phi_0 = (2n-1)/4$ with $n$ an integer, by exhibiting a discontinuous jump. The jump amplitude also decreases with increased system size, indicating the first order quantum phase transition will be changed into a second order one in the thermodynamic limit.  The $\Phi_0$ periodicity in the pairing order parameter renders that the supercurrent does not develop fully the $\Phi_0/2$ periodicity,  in a mathematically rigorous sense, regardless of the system size and electron filling. The reason lies in the fact that for a pure $d_{x^2-y^2}$-wave superconductor, low energy quasiparticle states can be populated by the Doppler shift rapidly along the nodal direction, for which  the orientation dependent coherence length is divergent.
As a hallmark of superconducting state, one can see clearly that flux-induced current has a different sign 
from that of the normal state  in the region approximately from 
$\Phi/\Phi_0=n-1/4$ to $\Phi/\Phi_0=n+1/4$ (compare Fig.~\ref{FIG:SuperCurrent1}(b1-b2) 
with Fig.~\ref{FIG:NormalCurrent}).   
At the half filling, the particle hole symmetry holds, which makes a large level spacing at the fermi energy when the circumference of the cylinder is small. Therefore, the supercurrent exhibits an activation-like behavior for the magnetic flux close to $\Phi/\Phi_0 = n$, and is very different from that at $\Phi/\Phi_0= n/2$. It explains why the $\Phi_0$ periodicity of supercurrent is pronounced in this specific case. 
Furthermore, it is not difficult to anticipate that, {\em as a canonical mesoscopic effect}, the breaking of  $hc/2e$ periodicity will be even more stronger for a smaller zero-field $d$-wave pair potential, and therefore a larger momentum averaged 
superconducting length $\xi = \langle \hbar v_{\mathbf{k}}/\pi \Delta_\mathbf{k}\rangle_{\text{FS}}$, where $v_{\mathbf{k}}$ is the quasiparticle velocity.  
We notice that,  in the region $\Phi/\Phi_0\in[n-1/4,n+1/4]$, the flux dependence of the supercurrent in a $d$-wave superconducting cylinder is different from that in a square $d$-wave superconducting loop~\cite{FLoder:2008}, where a zig-zag feature was obtained. We argue that in the geometry considered in Ref.~\onlinecite{FLoder:2008},  the elastic scattering from hard-wall boundaries of a mesoscale system populates a significant number of the lower energy states, which 
plays an important role in the flux-dependent current.
Naturally, the increase of the cylinder circumference is one way to reduce the activation behavior, and therefore reducing the $\Phi/\Phi_0$ component in the Fourier spectrum of supercurrent. 
%However,  this way is not so efficient solely because of the existence of nodal quasiparticle in gapless %$d$-pairing state. 
Alternatively, when the electronic filling factor is tuned away from the half filling, the particle-hole symmetry is broken and the Fermi surface becomes more isotropic. The activation-like behavior in the supercurrent is replaced by a more linear-like behavior,  similar to the behavior exhibiting 
at $\Phi/\Phi_0=n/2$. In addition, the current peaks at $\Phi/\Phi_0=(2n-1)/4$ becomes more symmetrized about $I=0$ axis, tending to restore the more of $\Phi_0/2$ perodicity in supercurrent.
Figure~\ref{FIG:SuperCurrent2} shows the flux dependence of the 
$d$-wave order parameter and persistent current for various electron filling and system size but with $t^{\prime}=-0.2$.
When a finite next nearest-neighbor hopping integral is introduced, the particle-hole symmetry is broken at the outset for the hole doped (i.e., $n_e\leq 1$) system, where the chemical potential is not zero. 
In this case, the level repulsion at the Fermi energy is weakened for flux close to $\Phi/\Phi_0 = n$ even for a small cylinder circumference, and the activation behavior in the current does not show up. Consequently, the $\Phi_0$ component in the current spectrum is dramatically decreased, which makes the total current looks to have $\Phi_0/2$ periodicity completely. We note that when the zero-field averaged superconducting coherence is not so small in comparison to the cylinder circumference, the restoration is always incomplete, due to the mesoscopic effect.

\begin{figure}[t]
%%\centerline{\psfig{filefig2.eps,width=8cm}}
\includegraphics[width=8cm, angle=0]{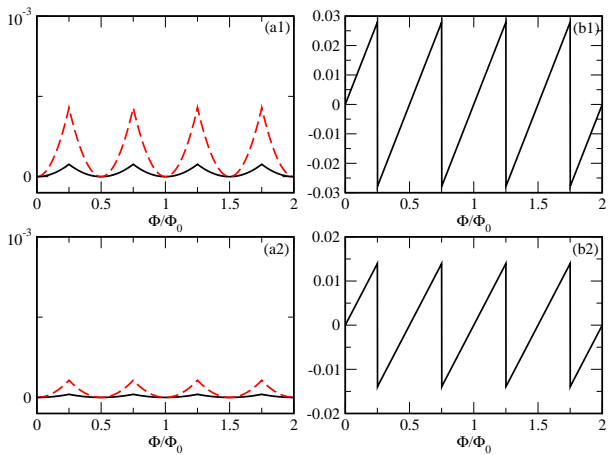}
\caption{(Color online)  Flux dependence of the $d$-wave order parameter (a1-a2) and persistent current (b1-b2) for a $d_{x^2-y^2}+id_{xy}$-wave superconducting cylinder of zie $N_L=40^2$ (a1-b1) and $N_{L}=80^2$ (a2-b2), with $n_e=1.0$. In (a1-a2),  the relative amplitude 
of the respective $d_{x^2-y^2}$ (solid line) and $d_{xy}$ (dashed line) components are plotted.
Here $t^{\prime}=0$ and the other parameter values are defined in the main text.
} \label{FIG:SuperCurrent3}
\end{figure}

To understand better the magnetic flux periodicity in unconventional superconductors, we turn to consider a $d_{x^2-y^2}+id_{xy}$ pairing state by taking the pairing strength in the $id_{xy}$ channel as $V_{xy}=3.0$. Now the quasiparticle excitions are gapfull. In Fig.~\ref{FIG:SuperCurrent3}, we show the flux dependence of the $d$-wave order parameter and supercurrent in a hollow $d_{x^2-y^2}+id_{xy}$-wave superconducting cylinder. Noticeably, even in the presence of the particle-hole symmetry and for the same system size as the case of a pure $d_{x^2-y^2}$ pairing state,  both the $d_{x^2-y^2}$ and $d_{xy}$ components of order parameter have the periodicity of $\Phi_0/2=hc/2e$. In particular, one case see that the evolution of two components of order parameter is continuous when $\Phi/\Phi_0$ crosses $(2n-1)/4$, indicating that the flux-induced first order quantum phase transition is unique to a cylinder formed by unconventional superconductors with nodal quasiparticles. In the present case, the magnetic $hc/2e$ periodicity in the current is complete. The $hc/2e$ periodicity is set in as long as the cylinder circumference is much larger than the superconducting coherence length. We point out (but do not show) that the flux dependence of order parameter and supercurrent in a hollow $s$-wave superconducting cylinder exhibit similar behavior to the case of $d_{x^2-y^2}+id_{xy}$ pairing state. 

One remark is in order: Our calculations have shown that the magnitude of the order parameter can be enhanced in the presence of magnetic flux throughout the whole period of $\Phi_0$ (see e.g., Fig.~\ref{FIG:SuperCurrent1}(a2) and Fig.~\ref{FIG:SuperCurrent3}(a1)-(a2)). This is the characteristic of all superconductors (including $s$-wave case) with short coherence length, which is much smaller than the cylinder circumference. We have calculated the case of a pure $d_{x^2-y^2}$-wave superconductor 
 and a conventional  $s$-wave superconductor both with a pairing interaction equal to $1$,  and found the flux dependence of the order parameter similar to that shown in Fig.~2 in Ref.~\onlinecite{JXZhu:1994} or Fig.~6 in Ref.~\onlinecite{KCzajka:2005}, which again is a mesoscopic effect.

In conclusion, we have studied the flux dependence of order parameter and supercurrent in a hollow 
$d$-wave superconducting cylinder. For a pure $d_{x^2-y^2}$ pairing state, we find a $hc/e$ periodicity of order parameter due to the existence of nodal quasiparticle states, and an associated quantum phase transition. When the particle hole symmetry holds in the normal state band structure, there is a noticeable component of $hc/e$ in the supercurrent spectrum when the cylinder circumference in the mesoscopic regime. However, in addtion to the increase of system size, this component can be suppressed more effectively by avoiding the particle hole symmetry point through the tuning of electron filling and band structure. By studying the case of a $d_{x^2-y^2}+id_{xy}$ pairing state, where the quasiparticle excitations are gapfull, we verify that the peculiar $hc/e$ magnetic flux periodicity only happens to unconventional superconductors with nodal structure.

{\bf Acknowledgments:} 
We thank Y.-Z. Wu for a technical help and T. Kopp for communication.
This work was carried out under the auspices of the National Nuclear Security Administration of the 
U.S. DOE at LANL under Contract No. DE-AC52-06NA25396, the LANL LDRD Programs, and the U.S. DOE Office of Science.

\end{document}